\documentclass[runningheads]{llncs}

\usepackage[springer]{definition}

\usepackage[misc]{ifsym}
\definecolor{sblue}{HTML}{02BCD4}
\definecolor{sred}{HTML}{F44436}
\definecolor{spink}{HTML}{E91E62}
\definecolor{sgreen}{HTML}{8BC34A}
\definecolor{spurple}{HTML}{3F51B5}
\definecolor{slightgreen}{HTML}{CCDE3A}
\definecolor{sorange}{HTML}{FE9800}
\definecolor{sgolden}{HTML}{FFC108}

\newcommand{\ie}[0]{{i.e.}}

\begin{document}
\annotator{yanqiao}{red}
\annotator{carl}{cyan}

\newcommand{\hejie}[1]{{\color{purple}[HJ: #1]}}
\title{Interpretable Graph Neural Networks for Connectome-Based Brain Disorder Analysis}

\titlerunning{Interpretable GNNs for Connectome-Based Brain Disorder Analysis}

\author{Hejie Cui\inst{1}\and 
Wei Dai\inst{1} \and
Yanqiao Zhu\inst{2} \and
Xiaoxiao Li\inst{3} \and \\
Lifang He\inst{4} \and
Carl Yang\inst{1}\textsuperscript{(\Letter)}}
\authorrunning{H. Cui et al.}

\institute{
\textsuperscript{1} Emory University \quad \textsuperscript{2} University of California, Los Angeles\\
\textsuperscript{3} The University of British Columbia \quad \textsuperscript{4} Lehigh University\\
\email{j.carlyang@emory.edu}
}

\maketitle
\begin{abstract}
Human brains lie at the core of complex neurobiological systems, where the neurons, circuits, and subsystems interact in enigmatic ways. Understanding the structural and functional mechanisms of the brain has long been an intriguing pursuit for neuroscience research and clinical disorder therapy. Mapping the connections of the human brain as a network is one of the most pervasive paradigms in neuroscience. Graph Neural Networks (GNNs) have recently emerged as a potential method for modeling complex network data. Deep models, on the other hand, have low interpretability, which prevents their usage in decision-critical contexts like healthcare. To bridge this gap, we propose an interpretable framework to analyze disorder-specific Regions of Interest (ROIs) and prominent connections. The proposed framework consists of two modules: a brain-network-oriented backbone model for disease prediction and a globally shared explanation generator that highlights disorder-specific biomarkers including salient ROIs and important connections. We conduct experiments on three real-world datasets of brain disorders. The results verify that our framework can obtain outstanding performance and also identify meaningful biomarkers. All code for this work is available at \url{https://github.com/HennyJie/IBGNN.git}.

\end{abstract}

\section{Introduction}
Brain networks (a.k.a the connectome) are complex graphs with anatomic regions represented as nodes and connectivities between the regions as links. Interpretable models on brain networks for disorder analysis are vital for understanding the biological functions of neural systems, which can facilitate early diagnosis of neurological disorders and neuroscience research \cite{maartensson2018stability}. Previous work on brain networks has studied models from shallow to deep, such as graph kernels \cite{jie2016sub}, tensor factorizations \cite{Liu:2018ty}, and convolutional neural networks \cite{BrainNetCNN, li2020braingnn, brainnetworktransformer}.

Recently, Graph Neural Networks (GNNs) attract broad interest due to their established power for analyzing graph-structured data \cite{kipf2016semi, Velickovic:2018we}. Compared with shallow models, GNNs are suitable for brain network analysis with universal expressiveness to capture the sophisticated connectome structures \cite{maron2018invariant, cui2022braingb, zhu2022joint, yang2022data}. However, GNNs as a family of deep models are prone to overfitting and lack transparency in predictions, preventing their usage in decision-critical areas like disorder analysis. 
Although several methods have been proposed for GNN explanation \cite{ying2019gnnexplainer, luo2020parameterized, DBLP:conf/nips/VuT20}, most of them focus on node-level prediction tasks and will produce a unique explanation for each subject when applied to graph-level tasks. However, for graph-level connectome-based disorder analysis, it is recognized that subjects having the same disorder share similar brain network patterns \cite{kan2022fbnetgen}, which means disorder-specific explanations across instances are preferable. Moreover, brain networks have unique properties such that directly applying vanilla GNN models will obtain suboptimal performance. 

In this work, we propose an interpretable GNN framework to investigate disease-specific patterns that are common across the group and robust to individual image quality. Meanwhile, the group-level interpretation can be combined with subject-specific brain networks for different levels of interpretation. As shown in Fig.~\ref{fig:framework}, it is composed of two modules: a backbone model IBGNN which adapts a message passing GNN designed for connectome-based disease prediction and an explanation generator that learns a globally shared mask to highlight disorder-specific biomarkers including salient Regions of Interest (ROIs) and important connections.
Furthermore, we combine the two modules by enhancing the original brain networks with the learned explanation mask and further tune the backbone model. The resulting model, which we term IBGNN+ for brevity, produces predictions and interpretations simultaneously.

Through experiments on three real-world brain disorder datasets (i.e. HIV, BP, and PPMI), we show our backbone model performs well across brain networks constructed from different neuroimaging modalities.
Also, it is demonstrated that the explanation generator can reveal disorder-specific biomarkers coinciding with neuroscience findings.
Last, we show that the combination of explanation generator and backbone model can further boost disorder prediction performance.

\section{The Proposed Model}

\paragraph{Problem definition.}
The input to the proposed framework is a set of $N$ weighted brain networks. For each network \(G = (V, E, \bm{W})\), \(V = \{v_i\}_{i = 1}^M\) is the node set of size \(M\) defined by the Regions Of Interest (ROIs) on a specific brain parcellation \cite{figley2017probabilistic, shirer2012decoding}, with each $v_i$ initialized with the node feature $\bm{x}_i$, \(E = V \times V\) is the edge set of brain connectome, and \(\bm{W} \in \mathbb{R}^{M \times M}\) is the weighted adjacency matrix describing the connection strengths between ROIs. The model outputs a brain disorder prediction \(\hat{y}_n\) for each subject $n$ and learns a disorder-specific interpretation matrix \(\bm{M} \in \mathbb{R}^{M\times M}\) that is shared across all subjects to highlight the disorder-specific biomarkers.

\begin{figure}[t]
	\centering
	\includegraphics[width=\linewidth]{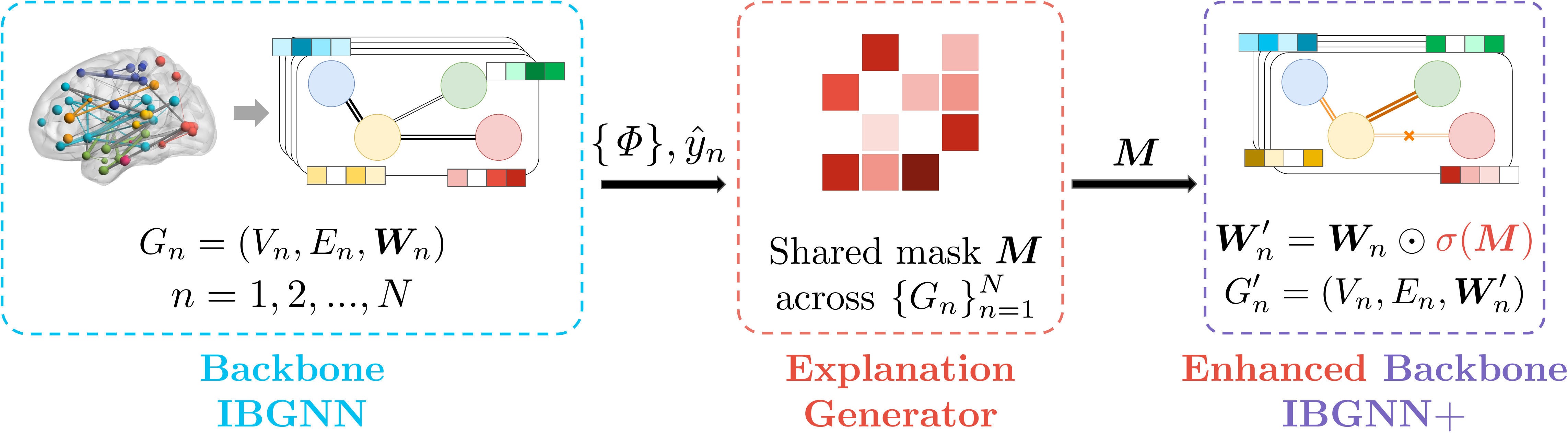}
	\caption{An overview of our proposed framework. The backbone model is firstly trained on the original data. Then, the explanation generator learns a globally shared mask across subjects. Finally, we enhance the backbone by applying the learned explanation mask and fine-tune the whole model.}
	\label{fig:framework}
\end{figure}

\paragraph{The backbone model IBGNN.}
\label{sec:predict}
\label{sec:predict_mp}
Edge weights in brain networks are often determined by the signal correlation between brain areas, which may have both positive and negative values, and thus cannot be handled correctly by conventional GNNs. To avoid this issue and better utilize edge weights in the GNN model, we design an edge-weight-aware message passing mechanism specifically for brain networks. Specifically, we first construct a message vector \(\bm{m}_{ij} \in \mathbb{R}^{D}\) by concatenating embeddings of a node \(v_i\) and its neighbor \(v_j\), and the edge weight \(w_{ij}\):
\begin{equation}
	\bm{m}_{ij}^{(l)} = \text{MLP}_1 \left( \left[ \bm{h}_i^{(l)};\, \bm{h}_j^{(l)};\, w_{ij} \right] \right),
\end{equation}
where \(l\) is the index of the GNN layer. Then, for each node $v_i$, we aggregate messages from all its neighbors $\mathcal{N}_i$ with the following propagation rule:
\begin{equation}
	\bm{h}_i^{(l)} = \xi \left( \sum\nolimits_{v_j \in \mathcal{N}_i \cup \{v_i\}} \bm{m}_{ij}^{(l - 1)} \right),
\end{equation}
where \(\xi\) is a non-linear activation function such as \(\operatorname{ReLU}\), and $\bm{h}_i^{(0)}$ is initialized with node feature $\bm{x}_i$ reflecting the connectivity information in brain networks \cite{cui2021positional}. 
After stacking $L$ layers, a readout function summarizing all node embeddings is employed to obtain a graph-level embedding $\bm{g}$. Formally, we instantiate this function with another Multi-Layer Perceptron (MLP) and residual connections:
\begin{equation}
	\bm{z} = \sum\nolimits_{i \in V} \bm{h}_{i}^{(L)}, \qquad \bm{g} = \text{MLP}_2 (\bm{z}) + \bm{z}.
\end{equation}
This backbone model IBGNN can be trained with the conventional supervised cross-entropy objective towards ground-truth disorder prediction, defined as 
\begin{equation}
\mathcal{L}_{\textsc{clf}}= - \frac{1}{N} \sum\nolimits_{n=1}^{N} 
(y_{n} \log \left(\hat{y}_{n}\right) + (1 - y_{n}) \log \left(1 - \hat{y}_{n}\right)).
\end{equation}

\paragraph{The globally shared explanation generator.}
\label{sec:explainer}
A general paradigm to generate explanations for GNNs is to find an explanation graph $G^{\prime}$ that has the maximum agreement with the label distribution on the original graph $G = (V, E, \bm{W})$, where $G^{\prime}$ can be a subgraph \cite{ying2019gnnexplainer} or other variations of $G$ \cite{luo2020parameterized, yuan2020explainability}. However, these explanation methods for GNNs mostly work on node-level prediction tasks and will produce a unique explanation graph for each subject when applied to graph-level tasks.
On the other hand, directly using attention weights in some attention-based GNN models \cite{Velickovic:2018we,yun2019graph} as explanations is known to be problematic \cite{DBLP:conf/naacl/JainW19, bai2021attentions}. 
Note that brain networks have some unique properties. For example, the node number and order are fixed under a given atlas. Also, brain networks assume that subjects with the same brain disorder have similar brain connection patterns. Therefore, a globally shared explanation graph $G^{\prime}$ capture common patterns for specific disorders at the group level is preferable.

In this work, we propose to learn a globally shared edge mask $\bm{M}\in \mathbb{R}^{M \times M}$ that is applied to all brain network subjects in a dataset. 
Specifically, we maximize the agreement between the predictions $\hat{y}$ on the original graph $G$ and $\hat{y}^{\prime}$ on an explanation graph $G^{\prime} = (V, E, \bm{W}^{\prime})$ induced by a masking matrix \(\bm{M}\), where $\bm {W}^{\prime} = \bm W \odot \sigma(\bm M)$, $\odot$ denotes element-wise multiplication, and $\sigma$ denotes the sigmoid function.
Formally this objective is implemented as a cross-entropy loss: 
\begin{equation}
\mathcal{L}_\textsc{mask} =  -\frac{1}{N}\sum\nolimits_{n=1}^{N}\sum\nolimits_{c=1}^{C} \mathds{1} [\hat{y}_{n}=c] \log P_{\Phi}\left(\hat{y}_{n}^{\prime}=\hat{y}_{n} \mid  G_{n}^{\prime} \right),
\end{equation} 
where $\sum_{n=1}^{N} P_{\Phi}\left(\hat{y}_{n}^{\prime}=\hat{y}_{n} \mid G_{n}^{\prime} \right)$ represents the conditional probability that the backbone model $\Phi$'s prediction $\hat{y}_{n}^{\prime}$ on the masked graph $G_{n}^{\prime}$ is consistent with the prediction $\hat{y}_n$ on the original graph $G_n$, C is the number of possible prediction labels. 
Besides, following the practice in GNNExplainer \cite{ying2019gnnexplainer}, we further apply two regularization terms \(\mathcal{L}_\textsc{sps}\) and \(\mathcal{L}_\textsc{ent}\) to encourage the compactness of the explanation and the discreteness of the mask values, respectively: 
\begin{equation}
    \mathcal{L}_\textsc{sps}=\sum\nolimits_{i, j} \bm M_{i, j}, \qquad \mathcal{L}_\textsc{ent} = -(\bm M \log (\bm M)+(1-\bm M) \log (1-\bm M)).
\end{equation}
The final training objective is given as:
\begin{equation}
\mathcal{L} = \mathcal{L}_\textsc{clf} + \alpha \mathcal{L}_\textsc{mask} + \beta \mathcal{L}_\textsc{sps} + \gamma \mathcal{L}_\textsc{ent}, 
\end{equation}
where $\alpha$, $\beta$ and $\gamma$ scale the numerical value of each loss item to the same order of magnitude to balance their influence. Our explanation generator will generate a globally shared edge mask that can be used for all testing graphs to investigate neurological biomarkers and highlight disorder-specific salient connections.

\paragraph{Enhancing the backbone with the learned explanations.}
The learned explanation mask can further improve the disorder prediction considering that raw brain networked data inevitably contain random noise.
Specifically, we enhance the original backbone by applying essential disorder-specific signals.
We note that this strategy is compatible with any backbone model, not limited to our proposed IBGNN.
We combined the aforementioned two modules so that predictions and interpretations are produced in a closed-loop for brain disorder analysis.
We term the enhanced model by IBGNN+ hereafter.

The whole training pipeline is summarized in Fig.~\ref{fig:framework}. The original brain networks are firstly input to train the backbone model. Then, a globally shared explanation mask is learned based on the backbone model $\Phi$ and prediction $\hat{y}_{n}$. Finally, we enhance the backbone model by highlighting salient ROIs and important connections on the raw data and tune the backbone model again.
\section{Experiments}

\paragraph{Dataset acquisition and preprocessing.}
We evaluate our framework using three real-world neuroimaging datasets of different modalities.
Specifically, groups in each dataset have balanced age and gender portions and are collected with the same image acquisition procedure. 
\begin{itemize}[nosep]
    \item \textit{Human Immunodeficiency Virus Infection (HIV): } This dataset is collected from Early HIV Infection Study at Northwestern University. It includes fMRI imaging of 70 subjects, 35 of which are early HIV patients, and the others are seronegative controls. We perform image preprocessing using the DPARSF\footnote{\url{http://rfmri.org/DPARSF/}} toolbox. The images are realigned to the first volume, followed by slice timing correction, normalization, spatial smoothness using an 8-mm Gaussian kernel, band-pass filtering (0.01-0.08 Hz), and linear trend removing of the time series. We focus on the 116 anatomical regions of interest (ROI), and extract a sequence of responses from them. Finally, brain networks with 90 cerebral regions are constructed, where each node represents a brain region and links are created based on correlations between different brain regions.
    \item \textit{Bipolar Disorder (BP): } This DTI imaging dataset is collected from 52 bipolar I subjects and 45 healthy controls. We use the FSL toolbox\footnote{\url{https://fsl.fmrib.ox.ac.uk/fsl/fslwiki/}} for preprocessing which includes distortion correction, noise filtering, and repetitive sampling from the distributions of principal diffusion directions for each voxel. Each subject is parcellated into 82 regions based on FreeSurfer-generated cortical/subcortical gray matter regions. 
    \item \textit{Parkinson's Progression Markers Initiative (PPMI): } This large-scale, publicly available dataset\footnote{\url{https://www.ppmi-info.org/}} is from a collaborative study\footnote{\url{https://www.michaeljfox.org/}} to improve PD therapeutics. We consider brain imaging in the DTI modality of 754 subjects, 596 of whom are Parkinson's disorder patients, and the rest 158 are healthy controls. The raw data are aligned using the FSL eddy-correct tool to correct head motion and eddy current distortions. The brain extraction tool (BET) from FSL is used to remove non-brain tissue. The skull-stripped images are linearly aligned and registered using Advanced Normalization Tools (ANTs\footnote{\url{http://stnava.github.io/ANTs/}}). 84 ROIs are parcellated from T1-weighted structural MRI using FreeSurfer\footnote{\url{https://surfer.nmr.mgh.harvard.edu/}} and the brain network connectivity is reconstructed using the deterministic 2nd-order Runge-Kutta (RK2) whole-brain tractography algorithm \cite{zhan2015comparison}.
\end{itemize}

\paragraph{Experimental settings.}
The proposed model is implemented using PyTorch 1.10.2 \cite{Paszke:2019vf} and PyTorch Geometric 2.0.3 \cite{Fey:2019wv}. A Quadro RTX 8000 GPU with 48GB of memory is used for our model training. Hyper-parameters are selected automatically with the open source AutoML toolkit NNI\footnote{\url{https://github.com/microsoft/nni}\label{nni}}.
We refer readers of interest to supplementary materials for implementation details. All reported results are averaged of ten-fold cross validation.

\paragraph{Baselines.}
We compare our proposed models, \ie, the backbone model IBGNN and the explanation enhanced IBGNN+, with competitors of both shallow and deep models. Shallow methods include M2E \cite{Liu:2018ty}, MIC \cite{Shao:2015ek}, MPCA \cite{Lu:2008cw}, and MK-SVM \cite{Dyrba:2015ci}, where the output graph-level embeddings are evaluated using logistic regression classifiers. We also include three representative deep graph models: GAT \cite{Velickovic:2019tu}, GCN \cite{kipf2016semi}, PNA \cite{corso2020principal} and two state-of-the-art deep models specifically design for brain networks: BrainNetCNN \cite{BrainNetCNN} and BrainGNN \cite{li2020braingnn}. 

\paragraph{Prediction performance.}
\begin{table}[t]
	\centering
	\caption{Experimental results (\%) on three datasets, where * denotes a significant improvement according to paired \textit{t}-test with $p=0.05$ compared with baselines. The best performances are in bold and the second runners are underlined.} 
	\resizebox{\linewidth}{!}{
	\begin{tabular}{cccccccccc}
	\toprule
	\multirow{2.5}{*}{Method} & \multicolumn{3}{c}{HIV} & \multicolumn{3}{c}{BP} & \multicolumn{3}{c}{PPMI} \\
	\cmidrule(lr){2-4} \cmidrule(lr){5-7} \cmidrule(lr){8-10}
	& Accuracy & F1 & AUC & Accuracy & F1 & AUC & Accuracy & F1 & AUC \\
	\midrule
	M2E   & 57.14{\tiny±19.17} & 53.71{\tiny±19.80} & 57.50{\tiny±18.71} & 52.56{\tiny±13.86} & 51.65{\tiny±13.38} & 52.42{\tiny±13.83} & 78.69{\tiny±1.78} & 45.81{\tiny±4.17} & 50.39{\tiny±2.59}\\
	MIC   & 54.29{\tiny±18.95} & 53.63{\tiny±19.44} & 55.42{\tiny±19.10} & 62.67{\tiny±20.92} & 63.00{\tiny±21.61} & 61.79{\tiny±21.74} & 79.11{\tiny±2.16} & 49.65{\tiny±5.10} & 52.39{\tiny±2.94} \\
	MPCA  & 67.14{\tiny±20.25} & 64.28{\tiny±23.47} & 69.17{\tiny±20.17} & 52.56{\tiny±13.12} & 50.43{\tiny±14.99} & 52.42{\tiny±13.69} & 79.15{\tiny±0.57} & 44.18{\tiny±0.18} & 50.00{\tiny±0.00} \\
	MK-SVM & 65.71{\tiny±7.00} & 62.08{\tiny±7.49} & 65.83{\tiny±7.41} & 57.00{\tiny±8.89} & 41.08{\tiny±13.44} & 53.75{\tiny±8.00}     & 79.15{\tiny±0.57} & 44.18{\tiny±0.18} & 50.00{\tiny±0.00} \\
	\midrule
	GCN   & 70.00{\tiny±12.51} & 68.35{\tiny±13.28} & 73.58{\tiny±9.49} & 55.56{\tiny±13.86} & 50.71{\tiny±11.75} & 61.55{\tiny±28.77}  & 78.55{\tiny±1.58} & 47.87{\tiny±4.40} & 59.43{\tiny±8.64} \\
	GAT   & 71.43{\tiny±11.66} & 69.79{\tiny±10.83} & 77.17{\tiny±9.42} & 63.34{\tiny±9.15} &  60.42{\tiny±7.56} & 67.07{\tiny±5.98}    & 79.02{\tiny±1.25} & 45.85{\tiny±3.16} & 64.40{\tiny±6.87} \\
	PNA & 57.14{\tiny±12.78} & 45.09{\tiny±19.62} & 57.14{\tiny±12.78} & 63.71{\tiny±11.34} & 55.54{\tiny±14.06} & 60.30{\tiny±11.89}   & 79.36{\tiny±1.84} & 51.76{\tiny±10.32} & 54.71{\tiny±6.77}\\
	BrainNetCNN &  69.24{\tiny±19.04}& 67.08{\tiny±11.11} & 72.09{\tiny±19.01} & 65.83{\tiny±20.64} & 64.74{\tiny±17.42} & 64.32{\tiny±13.72} & 55.20{\tiny±12.63} & \underline{55.45{\tiny±9.15}} & 52.54{\tiny±10.21}\\
	BrainGNN & 74.29{\tiny±12.10} & 73.49{\tiny±10.75} & 75.00{\tiny±10.56} & 68.00{\tiny±12.45} & 62.33{\tiny±13.01} & 74.20{\tiny±12.93} &69.17{\tiny±0.00}  & 44.19{\tiny±0.00} &45.26{\tiny±3.65}\\
	\midrule 
	\rowcolor{lightgray!20} IBGNN & \underline{82.14{\tiny±10.81}}\textsuperscript{*} & \underline{82.02{\tiny±10.86}}\textsuperscript{*} & \underline{86.86{\tiny±11.65}}\textsuperscript{*} & \underline{73.19{\tiny±12.20}} & \underline{72.87{\tiny±12.09}}\textsuperscript{*} & \underline{83.64{\tiny±9.61}}\textsuperscript{*} & \textbf{79.82{\tiny±1.47}} & 51.58{\tiny±4.66} & \underline{70.65{\tiny±6.55}}\textsuperscript{*} \\
	\rowcolor{lightgray!20} IBGNN+ & \textbf{84.29{\tiny±12.94}}\textsuperscript{*} & \textbf{83.86{\tiny±13.42}}\textsuperscript{*} & \textbf{88.57{\tiny±10.89}}\textsuperscript{*} & \textbf{76.33{\tiny±13.00}}\textsuperscript{*} & \textbf{76.13{\tiny±13.01}}\textsuperscript{*} & \textbf{84.61{\tiny±9.08}}\textsuperscript{*} & \underline{79.55{\tiny±1.67}} & \textbf{56.58{\tiny±7.43}} & \textbf{72.76{\tiny±6.73}}\textsuperscript{*} \\
	\bottomrule
	\end{tabular}}
	\label{tab:performance}
\end{table}
The overall results are presented in Table~\ref{tab:performance}. Both our proposed models yield impressive improvements over SOTA shallow and deep baselines. Compared with shallow models such as MK-SVM, our backbone model IBGNN outperforms them by large margins, with up to 11\% absolute improvements on BP. Besides, the effectiveness of our brain network-oriented design is supported by its superiority compared with other SOTA deep models. Moreover, the performance of the explanation enhanced model IBGNN+ can further increase the backbone by about 9.7\% relative improvements, which demonstrates that IBGNN+ effectively highlights the disorder-specific signals while also achieving the benefit of restraining random noises in individual graphs. 

\section{Interpretation Analysis}
\paragraph{Neural system mapping.}
The ROIs on brain networks can be partitioned into neural systems based on their structural and functional roles under a specific parcellation atlas, which facilitates the understanding of generated explanations from a neuroscience perspective. In this paper, we map the ROI nodes as defined on each dataset into eight commonly used neural systems, including Visual Network (VN), Auditory Network (AN), Bilateral Limbic Network (BLN), Default Mode Network (DMN), Somato-Motor Network (SMN), Subcortical Network (SN), Memory Network (MN), and Cognitive Control Network (CCN). 

\paragraph{Salient ROIs.}
\begin{figure}[t]
\centering
    \begin{minipage}{.32\textwidth}
      \centering
      \includegraphics[width=0.9\linewidth]{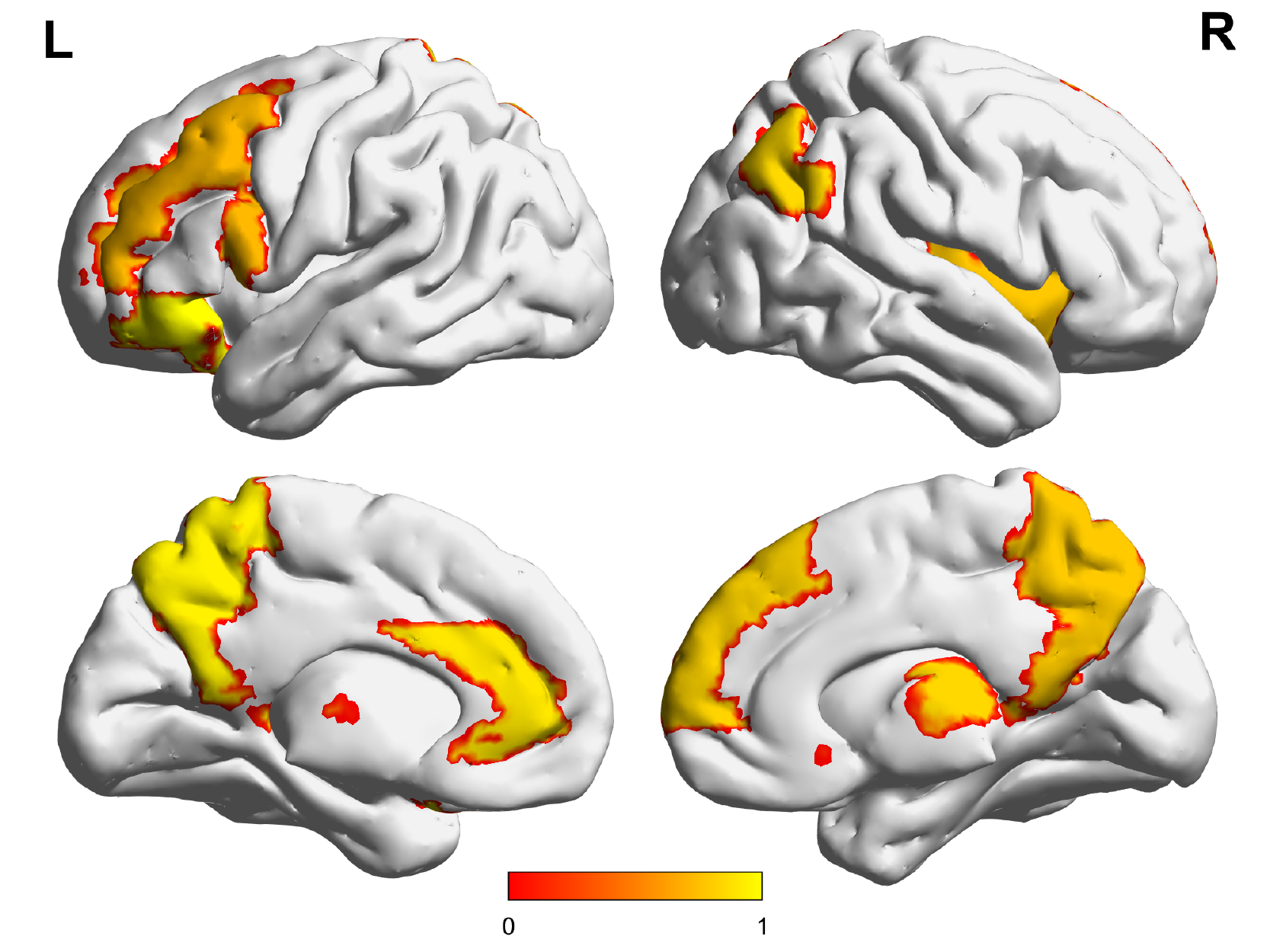}
      (a) HIV HC
      \label{fig:hiv_hc_nii}
    \end{minipage}%
    \begin{minipage}{.32\textwidth}
      \centering
      \includegraphics[width=0.9\linewidth]{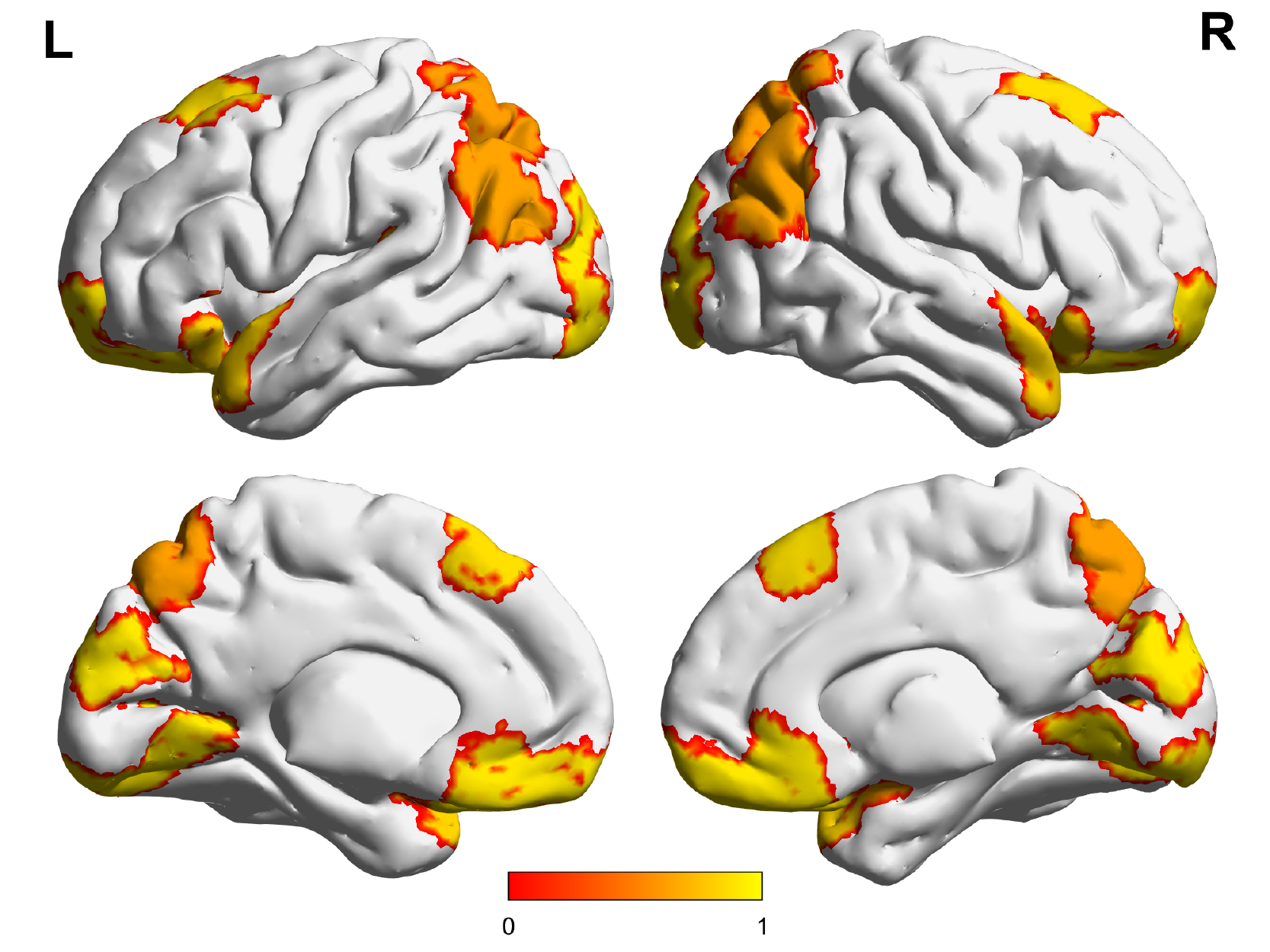}
      (c) BP HC
      \label{fig:bp_hc_nii}
    \end{minipage}
    \begin{minipage}{.32\textwidth}
      \centering
      \includegraphics[width=0.9\linewidth]{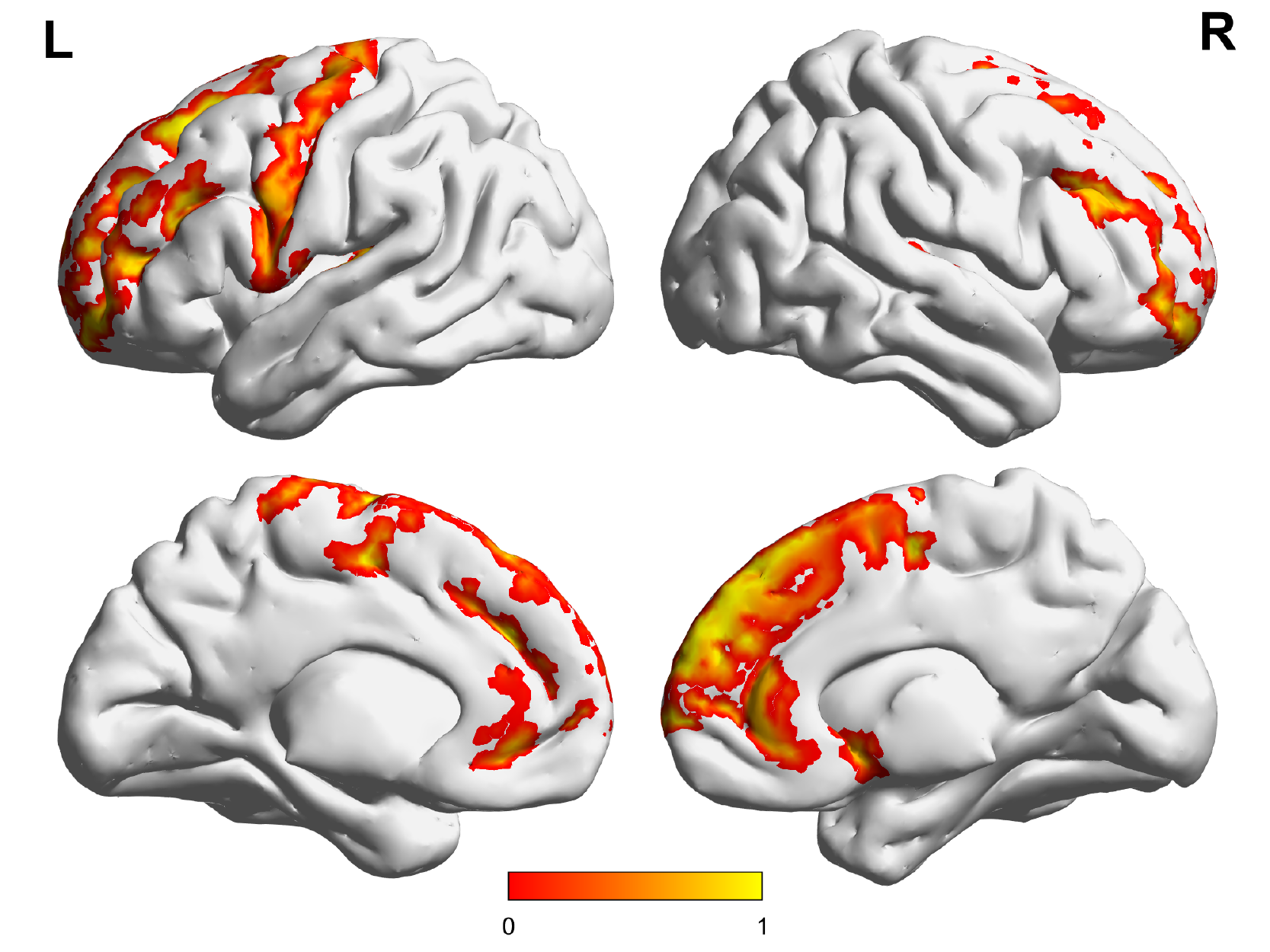}
      (e) PPMI HC
      \label{fig:ppmi_hc_nii}
    \end{minipage}
    \begin{minipage}{.32\textwidth}
      \centering
      \includegraphics[width=0.9\linewidth]{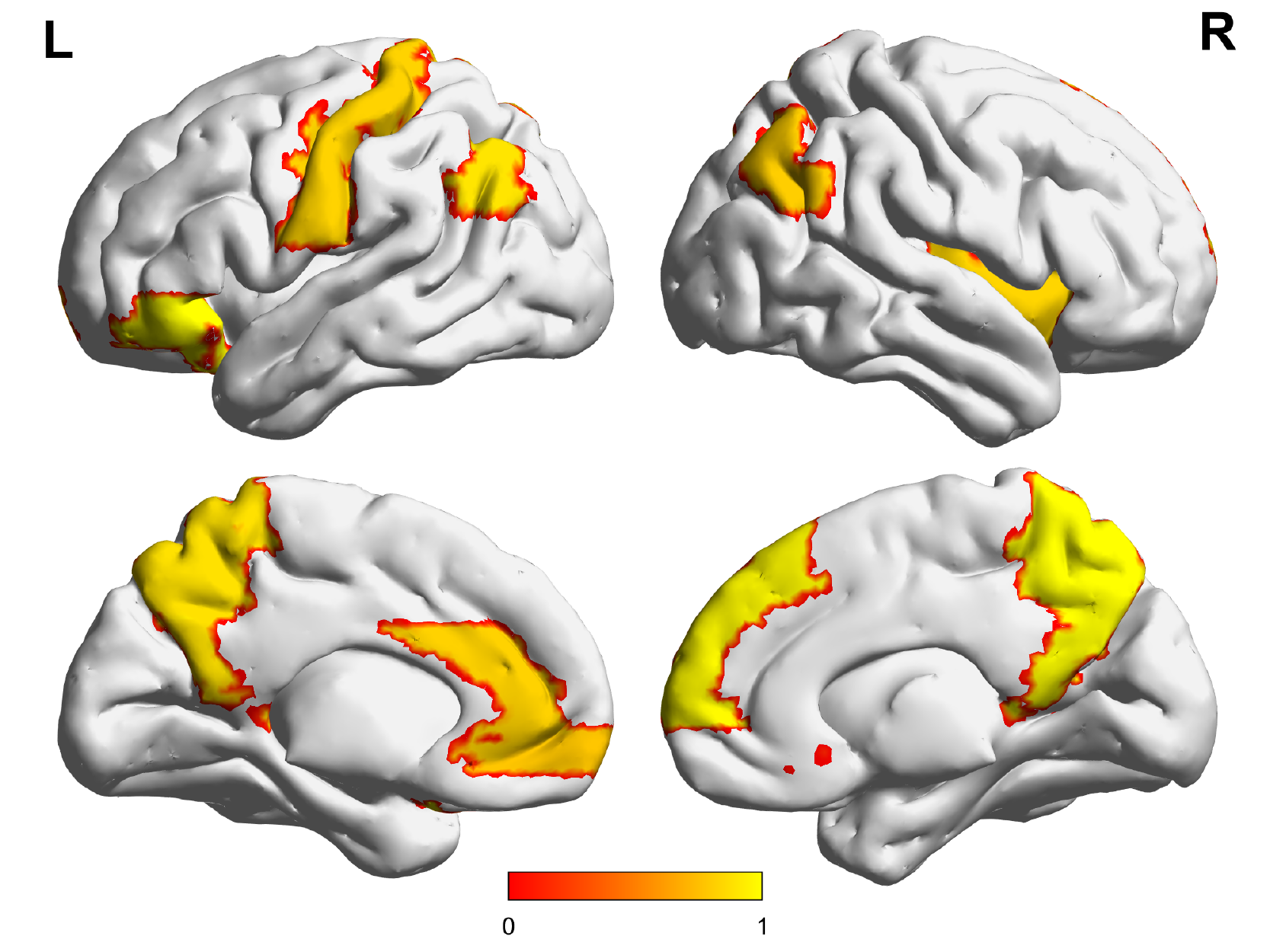}
      (b) HIV Patient
      \label{fig:hiv_patient_nii}
    \end{minipage}%
    \begin{minipage}{.32\textwidth}
      \centering
      \includegraphics[width=0.9\linewidth]{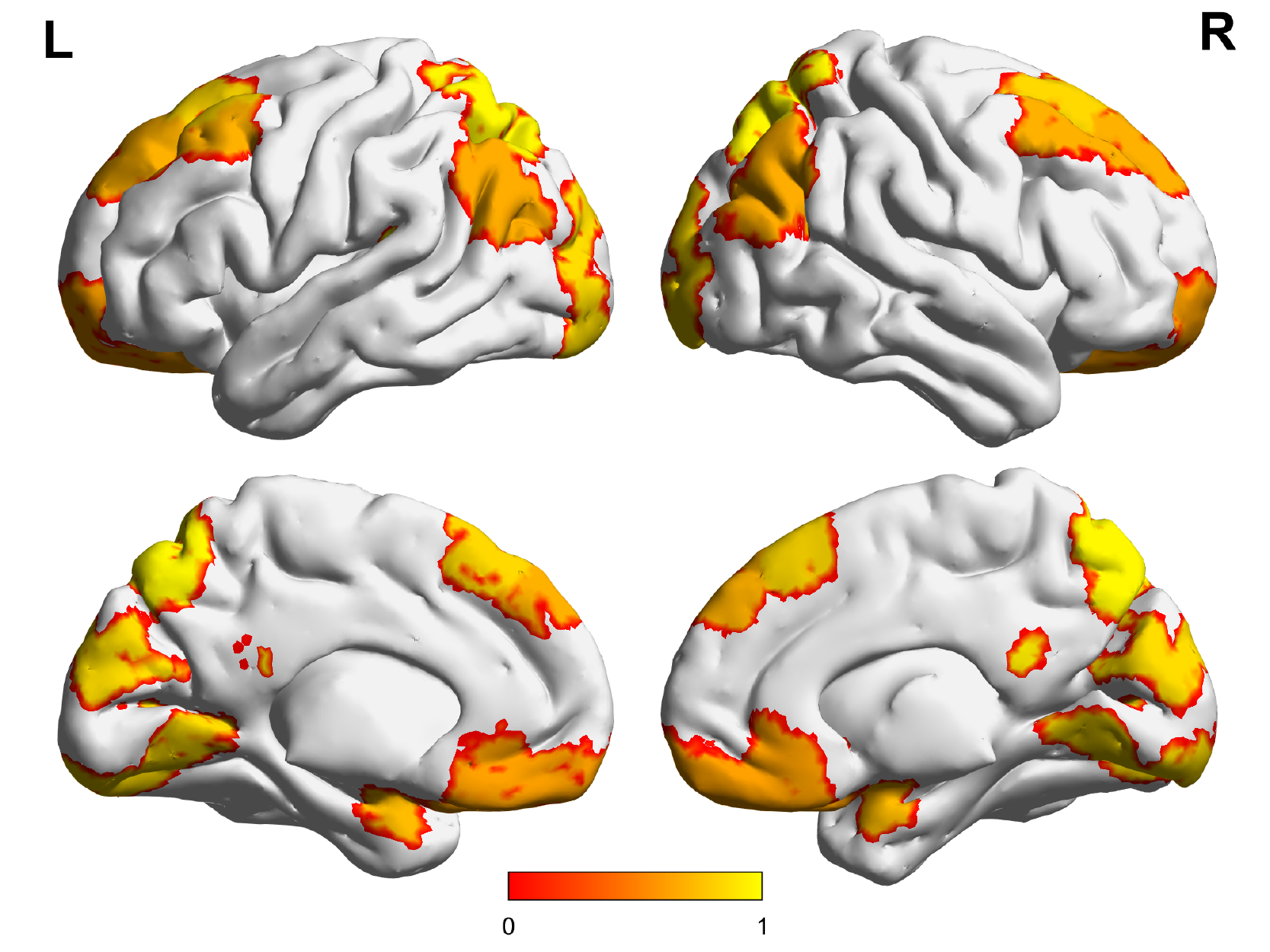}
      (d) BP Patient
      \label{fig:bp_patient_nii}
    \end{minipage}
    \begin{minipage}{.32\textwidth}
      \centering
      \includegraphics[width=0.9\linewidth]{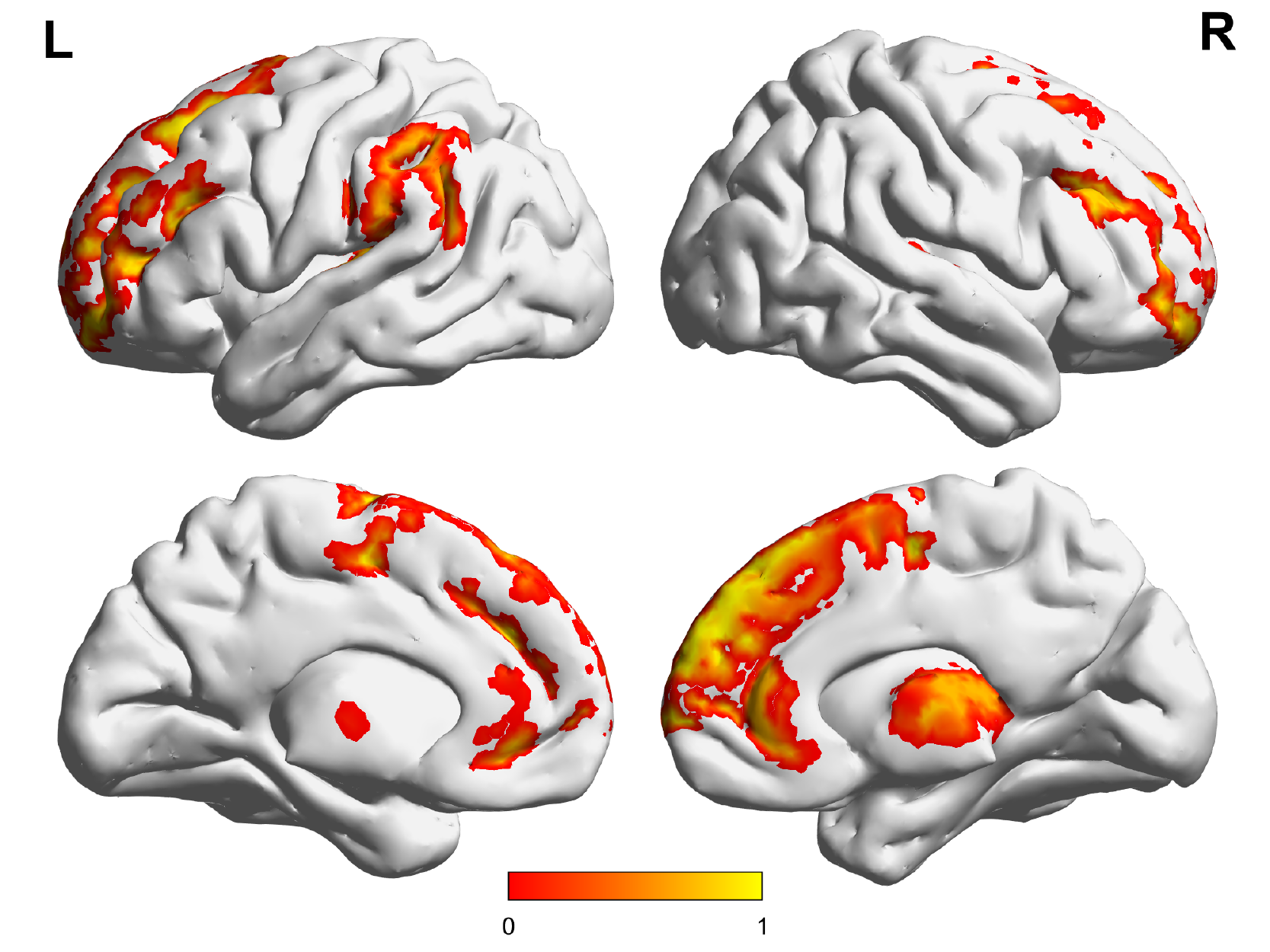}
      (f) PPMI Patient
      \label{fig:ppmi_patient_nii}
    \end{minipage}
\caption{\label{fig:nni_vis} Visualization of salient ROIs on the explanation enhanced brain connection networks for Health Control (HC) and Patient. The color of regions represents ROI's average importance in the given group. The bright-yellow color indicates a high score, while dark-red indicates a low score.}
\end{figure}
We provide both group-level and individual-level interpretations to understand which ROIs contribute most to the prediction of a specific disorder. On the group level, we rank the most salient ROIs on the learned explanation mask by calculating the sum of the edge weights connected to each node. Then on the individual level, we use the BrainNet Viewer \cite{xia2013brainnet} to plot the salient ROIs on the average brain connectivity graph enhanced by the learned explanation mask. 
For the HIV disease, anterior cingulate, paracingulate gyri, and inferior frontal gyrus are selected as salient ROIs. This complies with scientific findings that the regional homogeneity value of the anterior cingulate and paracingulate gyri are decreased \cite{ma2021hiv} and lower gray matter volumes are found in inferior frontal gyrus in HIV patients \cite{li2014structural}. The individual-level visualizations in Fig.~\ref{fig:nni_vis}(a)(b) show the difference between Health Control (HC) and HIV patients in those salient ROIs.
For the BP disease, secondary visual cortex and medial to superior temporal gyrus are selected as salient ROIs. This observation is in line with existing studies that visual processing abnormalities have been characterized in bipolar disorder patients \cite{reavis2020structural, o2014disturbances}, which is also confirmed in Fig.~\ref{fig:nni_vis}(c)(d).
For the PPMI disease, rostral middle frontal gyrus and superior frontal gyrus are selected as salient ROIs and Fig.~\ref{fig:nni_vis}(e)(f) display the difference. This is in accordance with MRI analysis revealing a significant decrease in PD patients in the rostral medial frontal gyrus and superior, middle, and inferior frontal gyri \cite{kendi2008altered}. 
All these observed salient ROIs can be potential biomarkers to identify brain disorders from each cohort.

\label{sec:vis}
\begin{figure}[t]
\centering
    \begin{minipage}{.33\textwidth}
      \centering
      \includegraphics[width=0.9\linewidth]{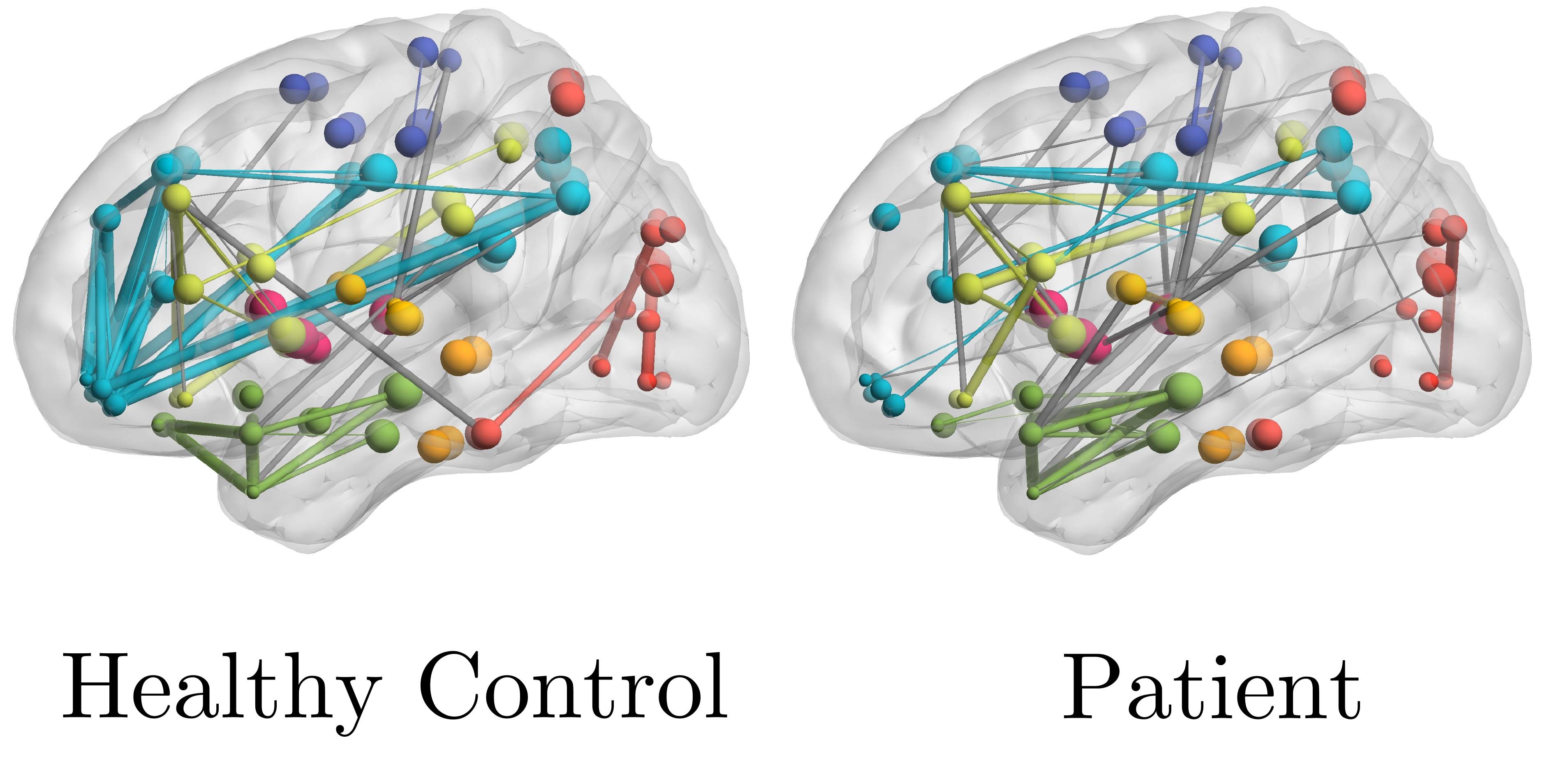}
      (a) HIV
      \label{fig:hiv_vis}
    \end{minipage}%
    \begin{minipage}{.33\textwidth}
      \centering
      \includegraphics[width=0.9\linewidth]{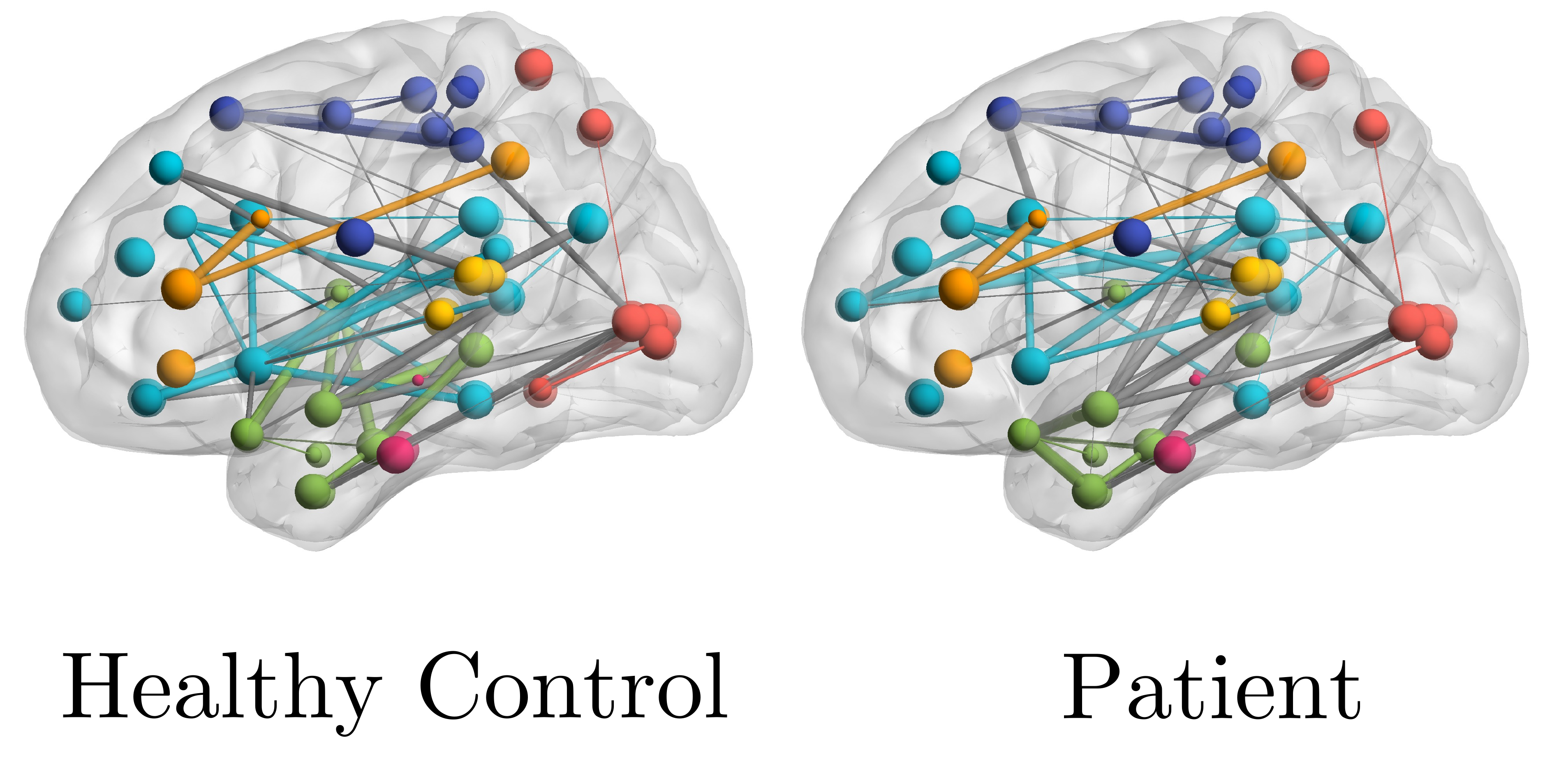}
      (b) BP
      \label{fig:bp_vis}
    \end{minipage}
    \begin{minipage}{.33\textwidth}
      \centering
      \includegraphics[width=0.9\linewidth]{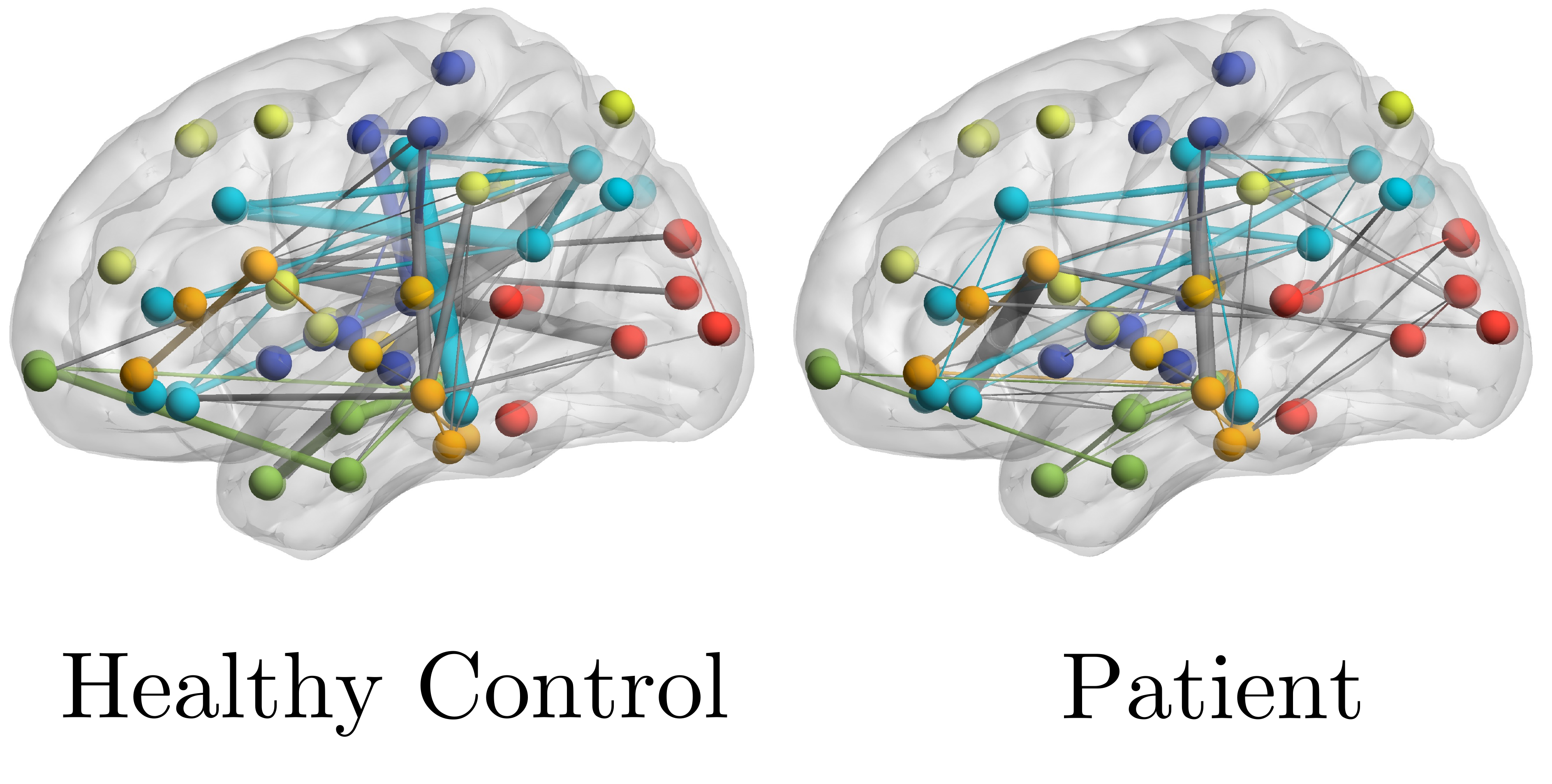}
      (c) PPMI
      \label{fig:ppmi_vis}
    \end{minipage}
\caption{\label{fig:vis} Visualization of important connections on the explanation enhanced brain connection network. Edges connecting nodes within the same neural system ({\color{sred}VN}, {\color{sgolden}AN}, {\color{sgreen}BLN}, {\color{sblue}DMN}, {\color{spurple}SMN}, {\color{spink}SN}, {\color{sorange}MN}, {\color{slightgreen}CCN}) are colored accordingly, while edges across different systems are colored gray. Edge width indicates its weight in the explanation graph.}
\end{figure}

\paragraph{Important connections.}
The globally shared explanation mask $\bm{M}$ provides interpretations of important connections. We obtain an explanation subgraph $G^{\prime}_s$ by taking the top 100 weighted edges from the masked $G^{\prime}$ with all other edges removed. The connection comparisons are shown in Fig.~\ref{fig:vis}, which helps identify connections related to specific disorders. 
For the HIV dataset, the explanation subgraph of patients excludes rich interactions within the DMN (colored {\color{sblue} blue}) system.  Also, interactions within the VN (colored {\color{sred}red}) system of patients are significantly less than those of HCs.
These patterns are consistent with the findings in earlier studies \cite{herting2015default, flannery2021hiv} that connectivity alterations within- and between-network DMN and VN may relate to known visual processing difficulties for HIV patients. 
For the BP dataset, compared with tight interactions within the BLN (colored {\color{sgreen} green}) system of the healthy control, the connections within BLN system of the patient subject are much sparser, which may signal pathological changes in this neural system. This observation is in line with previous studies \cite{das2020parietal}, which finds that the parietal lobe, one of the major lobes in the brain roughly located at the upper back area in the skull and is in charge of processing sensory information received from the outside world, is mainly related to Bipolar disorder attack. Since parietal lobe ROIs are contained in BLN under our parcellation, the connections missing within the BLN system in our visualization are consistent with existing clinical understanding.
For the PPMI dataset, the connectivity in the patient group decreases in the SMN (colored {\color{spurple} purple}) system, which integrates primary sensorimotor, premotor, and supplementary motor areas to facilitate voluntary movements. This observation confirms existing neuroimaging studies that have repeatedly shown disorder-related alteration in sensorimotor areas of Parkinson's patients \cite{caspers2021within}. Furthermore, individuals with PD have lower connectivity within the DMN (colored {\color{sblue} blue}) system compared with healthy controls, which is consistent with the cognition recession study on Parkinson's patients \cite{van2009dysfunction, tessitore2012default}.

\section{Conclusion}
In this work, we propose a novel interpretable GNN framework for connectome-based brain disorder analysis, which consists of a brain network-oriented GNN predictor and a globally shared explanation generator.
Experiments on real-world neuroimaging datasets show the superior prediction performance of both our backbone and the explanation enhanced models and validate the disorder-specific interpretations from the generated explanation mask. The limitation of the proposed framework might arise from the small size of neuroimaging datasets, which restraints the effectiveness and generalization ability of deep learning models.
A direct future direction based on this work is to utilize pre-training and transfer learning techniques to learn across datasets. This allows for the sharing of information and explanations across different cohorts, which could lead to a better understanding of cross-disorder commonalities.

\section*{Acknowledgement}
This research was partly supported by the internal funds and GPU servers provided by the Computer Science Department of Emory University and the University Research Committee of Emory University.
Xiaoxiao Li was supported by NSERC Discovery Grant (DGECR-2022-00430).
Lifang He was supported by ONR N00014-18-1-2009 and Lehigh's accelerator grant S00010293.

\bibliography{reference}
\clearpage
\end{document}